\documentclass[aps,prc,twocolumn,preprintnumbers,superscriptaddress,floatfix,showpacs]{revtex4-1}
\usepackage{epsfig}
\usepackage{dcolumn}
\usepackage{bm}
\usepackage{amssymb}
\usepackage{amsmath}
\usepackage[latin1]{inputenc}
\usepackage{color}

\begin{document}

\title{Viscous Effects on the Mapping of the Initial to Final State in Heavy Ion Collisions}

\author{Fernando G. Gardim}
\affiliation{Instituto de Ciência e Tecnologia, Universidade Federal de Alfenas, Cidade Universitária, 37715-400 Poços de Caldas, MG, Brazil}

\author{Jacquelyn Noronha-Hostler}
\affiliation{Department of Physics, Columbia University, New York, 10027, USA}
\affiliation{Instituto de F\'{i}sica, Universidade de S\~{a}o Paulo, C.P.
66318, 05315-970 S\~{a}o Paulo, SP, Brazil}

\author{Matthew Luzum}
\affiliation{McGill University,  3600 University Street, Montreal QC H3A 2TS, Canada}
\affiliation{Lawrence Berkeley National Laboratory, Berkeley, CA 94720, USA}
\affiliation{Departamento de F\'isica de Part\'iculas and IGFAE,
Universidade de Santiago de Compostela, E-15706 Santiago de
Compostela, Galicia-Spain}

\author{Frédérique Grassi}
\affiliation{Instituto de F\'{i}sica, Universidade de S\~{a}o Paulo, C.P.
66318, 05315-970 S\~{a}o Paulo, SP, Brazil}

\begin{abstract}
We investigate the correlation between various aspects of the initial geometry of heavy ion collisions at the Relativistic Heavy Ion Collider energies and the final anisotropic flow, using v-USPhydro, a 2+1 event-by-event viscous relativistic hydrodynamical model.  We test the extent of which shear and bulk viscosity affect the prediction of the final flow harmonics, $v_n$, from the initial eccentricities, $\varepsilon_n$.  We investigate in detail the flow harmonics $v_1$ through $v_5$ where we find that $v_1$, $v_4$, and $v_5$ are dependent on more complicated aspects of the initial geometry that are especially important for the description of peripheral collisions, including a non-linear dependence on eccentricities as well as a dependence on shorter-scale features of the initial density.  Furthermore, we compare our results to previous results from NeXSPheRIO, a 3+1 relativistic ideal hydrodynamical model that has a non-zero initial flow contribution, and find that the combined contribution from 3+1 dynamics and non-zero, fluctuating initial flow decreases the predictive ability of the initial eccentricities,  in particular for very peripheral collisions, but also disproportionately in central collisions. 
%
\end{abstract}

\maketitle

\section{Introduction}
Ultrarelativistic heavy ion collisions are essential in probing strongly interacting matter at high energy regimes. Relativistic hydrodynamics has been able to accurately describe a large amount of experimental data from these collisions  \cite{Heinz:2013th}. 
Predictions from ideal hydrodynamics describe quite well some observables, such as elliptic flow and higher flow harmonics \cite{Hama:2007dq,Huovinen:2001cy,Gardim:2012,Schenke:2010rr,Petersen:2010cw}, and di-hadron correlations \cite{Qian:2013nba}, which suggest that the Quark Gluon Plasma is a nearly perfect fluid. However, predictions are the most successful in the more central collisions (as compared to  peripheral ones) where the matter is the most dense and shear viscosity plays the smallest role. Furthermore, it has been recently shown that \cite{Noronha-Hostler:2013gga,Noronha-Hostler:2014dqa} bulk viscosity can compensate the effect of shear viscosity and can improve the fit to experimental data \cite{Rose:2014fba}, so it is important to include both viscous effects.  Thus, to improve the predictions and take into account dissipative effects, which are more important in smaller systems \cite{Heinz:2004ar}, viscous hydrodynamics is used in this paper within the event-by-event  code, v-USPhydro \cite{Noronha-Hostler:2013gga,Noronha-Hostler:2014dqa}.  However,  only small shear and bulk viscosities are considered due to the previous success of ideal hydrodynamics.

The study of strongly-coupled matter with hydrodynamics requires that one supply a set of initial conditions, then evolve them through ideal \cite{Takahashi:2009na,Petersen:2008dd,Holopainen:2010gz,Werner:2010aa,Qiu:2011iv}
or viscous~\cite{Noronha-Hostler:2013gga,Noronha-Hostler:2014dqa,Schenke:2010rr,Gale:2012rq,Niemi:2012ry,Qiu:2011hf}  hydrodynamics, and at the end, compute the particle emission. The particle distribution of each individual event reflects characteristics of the initial conditions, such as the energy density profile and the initial flow. In a non-central collision the averaged initial density profile presents an almond shape, which is commonly characterized by an elliptic eccentricity $\varepsilon_2$.  
Hydrodynamic evolution then converts this spatial asymmetry into an asymmetry in the final particle distribution, given by the elliptic flow $v_2$. Di-hadron azimuthal correlations data cannot be theoretically understood as coming from smooth density profiles, but rather in event-by-event hydrodynamics with fluctuations in the initial conditions \cite{Takahashi:2009na}. 
These fluctuations generate non-zero odd Fourier harmonics at mid-rapidity, for example a triangular anisotropy $v_3$ in the azimuthal particle distribution, as a consequence of an average triangular anisotropy in the initial density condition \cite{Alver:2010gr}. 

To construct more realistic models for early-time collision dynamics, and provide a more direct link between experimental data and the properties of initial conditions it is necessary to understand the anisotropic flow response to the initial state properties. Much effort has been made in that direction \cite{Alver:2010gr,Alver:2010dn,Petersen:2010cw,Schenke:2010rr,Qiu:2011iv,Qiu:2011hf,Luzum:2013yya,
Gardim:2011xv,Niemi:2012aj} -- where primarily only ideal hydrodynamics and no initial flow was used -- demonstrating that elliptic flow comes mainly from the elliptic shape and triangular flow comes from the initial triangularity $\varepsilon_3$, but quadrangular and pentagonal flow do not come only from quadrangularity and pentagularity \cite{Teaney:2010vd,Gardim:2011xv}, but also from combinations of $\varepsilon_2$ and $\varepsilon_3$, given by the cumulant expansion of the initial density profile \cite{Teaney:2010vd}. Directed flow $v_1$ seems to come from dipole asymmetry $\varepsilon_1$ \cite{Gardim:2011qn}, but it has not been well studied.

The goal of this work is to improve our understanding of the detailed relationship between the initial conditions and the final-measured observables.  We do this by solving event-by-event ideal and viscous hydrodynamics for Monte Carlo Glauber initial conditions, using both shear and bulk viscosity, as well as comparing the results to realistic initial conditions for ideal hydrodynamics, NeXSPheRIO.  Using this information, we quantitatively investigate the role of shear and bulk viscosity in the mapping of the initial state to the final state, as well  as various improved estimators of anisotropic flow $v_n$ for $n$=1--5.
%
%
%
%
%
%
%
%
%
%
\section{Mapping the hydrodynamic response}
In a purely hydrodynamic calculation of a relativistic heavy ion collision, particles are emitted independently from a fluid element, and all information is thus contained in the single-particle momentum distribution $dN/d^3p$ of each particle species in each event.   In principle correlations are present, but often they are negligible, and it has been shown that a purely hydrodynamic description can successfully describe a large range of experimental measurements.
%

The azimuthal dependence of the particle distribution is of particular interest, and can be usefully organized as a set of Fourier coefficients in the azimuthal angle $\phi_p$ of the outgoing particle momentum.  Here it will be convenient to write it as a complex Fourier series
\begin{equation}
\label{differential}
\frac {dN} {dy d^2p_T} = \frac 1 {2\pi} \frac {dN} {dy p_T dp_T} \sum_{n=-\infty}^\infty e^{in\phi_p} V_n(p_T, y) ,
\end{equation}
where $y$ is the particle rapidity and $p_T$ is the transverse momentum.  Note that, when written in this form, $V_0 = 1$ and $V_{-n} = V_n^*$.  Thus, all information about a single collision event is contained in a set of complex functions $V_n(p_T, y)$ with $n>0$ and the yield $dN/dydp_T(p_T, y)$ of each particle species.   Due to dynamical fluctuations, each event has a different particle distribution, and measured observables can be calculated as particular event-averaged functions of this distribution \cite{Luzum:2013yya}.

In this work we focus on momentum-integrated observables.   Integrating over $p_T$ and $y$, Eq.~\eqref{differential} becomes
\begin{equation}
\frac {dN} {d\phi_p} = \frac N {2\pi}  \sum_{n=-\infty}^\infty e^{in\phi_p} V_n ,
\end{equation}
with complex coefficients $V_n$ related to the common notation for the flow coefficients $v_n$ and event planes $\Psi_n$ as the magnitude and phase, respectively,
\begin{equation}
\label{vn}
V_n = v_n e^{i n \Psi_n} \equiv \frac 
{
\int d^3p \frac {dN} {d^3p} e^{in\phi_p}
} 
{
\int d^3p \frac {dN} {d^3p} 
} .
\end{equation}

In such a hydrodynamic calculation, the particle distribution is determined by the energy and momentum distribution $T^{\mu\nu}(\vec x)$ of the system at some initial time $\tau_0$, which is evolved forward in time and converted to a distribution of particles.  In principle, one can then further evolve the particle distribution function $f(\vec x, \vec p)$ with the Bolzmann equation.   Ultimately, each of these steps is completely deterministic, and so the above flow coefficients are a deterministic
functional of this initial condition
\begin{equation}
V_n = \mathcal{F}\left[ T^{\mu\nu}({\vec x}) \right] .
\end{equation}
Relativistic viscous hydrodynamics is a complicated set of nonlinear partial differential equations, and the distribution of particles emitted from a fluid element is also a complicated nonlinear function of the fluid properties.  Thus, it is not possible to write a simple, exact, analytic expression for this dependence on the initial condition.  However, it has been shown that, by using intuition and general arguments, one can find simple relations that give a very good approximation to these functionals \cite{Gardim:2011xv}.   With this information, we can in principle understand in detail what properties of the medium and initial condition are probed, which can consequently allow us to constrain these properties from measured data \cite{Retinskaya:2012ky,Retinskaya:2013gca}.  In addition, it can significantly reduce the amount of necessary computation \cite{Teaney:2012ke,Teaney:2013dta, Floerchinger:2014fta}, and allows for studies that would otherwise be unfeasible \cite{Luzum:2012wu}. 

The logic is as follows:  
We must find quantities derived from the initial condition that have the same general properties as the desired momentum-space observable,
and can therefore serve as an estimator of the final value after hydrodynamic evolution and freeze out.    For example, $V_n$ is a dimensionless quantity that behaves under azimuthal rotation of angle $\delta$ as
\begin{equation}
V_n \to V_n e^{in\delta} .
\end{equation}
In addition, it shouldn't depend on where one places the origin of the coordinate system $\vec x$ --- that is, a spatial translation of the initial energy distribution does not affect the momentum distribution of outgoing particles.

Any such ``estimator'' for $V_n$ is necessarily an approximation.  Thus, once a potential estimator $V_{est,n}$ is established, it is judged by how accurately it can predict $V_n$ on an event-by-event basis.  A given estimator may be an accurate estimation in  a particular specially-chosen event, but a poor estimation in other events.   It is therefore useful to define a measure of the quality of a proposed estimator that is preferably tested over a large and diverse set of realistic collision events.  The natural choice is the linear correlation coefficient between the flow vector $V_n$ and the estimator $V_{est, n}$ \cite{Gardim:2011xv}
\begin{equation}
\label{Q}
Q_n 
= \frac {{\rm Re}\left\langle V_n V_{est,n}^* \right\rangle} {\sqrt{\left\langle |V_{n}|^2 \right\rangle \left\langle |V_{est,n}|^2 \right\rangle}} .
\end{equation}
with angular brackets defined throughout this work as
\begin{equation}
\langle \ldots \rangle = \frac 1 {N_{events}} \sum_{events} \ldots
\end{equation}
Note that the numerator is an average of the scalar product of the two vectors, and $Q_n$ therefore represents the vector analogue of the Pearson correlation coefficient $r$ between the  vectors.  (We choose the notation $Q_n$ to avoid confusion with the quantity $r_n$ introduced in Ref.~\cite{Gardim:2012im}).
It is bounded by $\pm 1$, with a value of 1 obtained if and only if the estimator gives a perfect prediction of $V_n$ in every event.  
In general, larger values (i.e., values closer to 1) indicate a better linear correlation and therefore a better estimator.

One can also understand $Q_n$ in a different way.  In each event, one can define a 2-D error vector ${\cal E}_n$, defined by the difference between $V_n$ and the estimator $V_{est,n}$
\begin{equation}
{\cal E}_n \equiv V_n - V_{est,n}
\end{equation}
We can define the best estimator out of a set of candidates as the one that minimizes the average length of the error vector over a large set of events $\langle |{\cal E}_n|^2 \rangle$.

Most often, a proposed estimator is defined only up to an unknown overall (real) scaling coefficient $k$, 
\begin{equation}
V_{est,n} = k\ {\cal V}_{est,n} ,
\end{equation}
where ${\cal V}_{est,n}$ is an unscaled estimator. (As an explicit example, the most familiar relation is ${\cal V}_{est,n} = \varepsilon_n e^{in\Phi_n}$, with $k = v_n/\varepsilon_n$ the hydrodynamic response coefficient, to be discussed below).
%

In this case, the choice that minimizes $\langle |{\cal E}_n|^2 \rangle$ and gives the ``best estimator'' is \cite{Gardim:2011xv}
\begin{equation}
k = \frac 
{{\rm Re}\left\langle V_n {\cal V}_{est,n}^* \right\rangle}
{\left\langle |{\cal V}_{est,n}|^2 \right\rangle} ,
\end{equation}
and the following relation holds
\begin{equation}
\left\langle |{\cal E}_n|^2\right\rangle=\left\langle |V_n|^2\right\rangle-\left\langle |V_{est,n}|^2 \right\rangle.
\end{equation}
The quality $Q_n$ of the best estimator as defined in Eq.~\eqref{Q} can then also be written in terms of the error $\langle |{\cal E}_n|^2 \rangle$
\begin{equation}
Q^2_n \equiv 1 - \frac {\left\langle |{\cal E}_n|^2 \right\rangle} {\left\langle |V_n|^2 \right\rangle} 
= \frac {\left\langle |V_{est,n}|^2 \right\rangle} {\left\langle |V_n|^2 \right\rangle} .
\end{equation}
%

If a proposed estimator has multiple free parameters, one should choose their values to maximize $Q_n$.  For example, if we consider a linear sum of terms with coefficients $k_i$
\begin{equation}
V_{est,n} = \sum_{i=1}^N k_i {\cal V}_{n,i}  , 
\end{equation}
the best estimator is given by choosing coefficients $k_i$ to satisfy the set of linear equations
\begin{equation}
\label{linearequations}
\sum_{j=1}^N  k_j {\rm Re}\left\langle {\cal V}_{n,i} {\cal V}_{n,j}^*\right\rangle
= {\rm Re}\left\langle V_n {\cal V}_{n,i}^*\right\rangle
.
\end{equation}
\subsection{Characterizing the initial state}
\label{expansion}
The next step is to develop a system for choosing estimators and finding systematic improvements.

In hydrodynamics, rather than the components of the energy-momentum tensor $T^{\mu\nu}$ in a fixed frame, one typically describes the same information in terms of a fluid velocity $u^\mu$, the energy density in the fluid rest frame $\epsilon$, and the viscous tensor $\Pi^{\mu\nu}$, which is a small correction for a system in a fluid regime.  

If the system has an approximate boost invariance near mid-rapidity, and if the initial transverse flow and initial viscous tensor are negligible, the final momentum distribution depends only on the spatial distribution of energy density in the transverse plane $\epsilon(x,y)$.  It has been demonstrated that this is a good approximation \cite{Gardim:2011xv}, and is therefore a good starting point, from which corrections can be added.    Instead of energy density, one can equivalently consider entropy or enthalpy density.  In the following, we use the notation $\rho(\bf x)$ for a generic density in the transverse plane, ${\bf x} = (x,y)$.
%
%

Generically, hydrodynamics can be thought of as a description of long-wavelength modes.  Therefore, a reasonable starting point is to characterize the initial density as a set of modes that are ordered from large-scale to small-scale structure.  We do this by taking the (2-D) Fourier transform
\begin{equation}
\rho({\bf{k}}) = \int d^2x \rho({\bf{x}}) e^{i \bf{k}\cdot {\bf{x}}}
\end{equation}
and writing its logarithm as a Fourier series in azimuthal angle $\phi_k$ and a power series in the magnitude $k$ of the Fourier transform momentum \cite{Teaney:2010vd}
\begin{equation}
\label{cumulants}
W({\bf k}) \equiv \ln \rho({\bf{k}})  = \sum_{n=-\infty}^\infty \sum_{m\geq |n|} W_{n,m} k^m e^{-i n\phi_k } ,
\end{equation}
noting that nonzero terms only exist with $m\geq|n|$ and $(m-n)\in 2\mathbb{Z}$,
and that 
$W_{-n,m} \propto W_{n,m}^*$ 
so that only values for $n\geq0$
contain independent information.

If the function $\rho({\bf x})$ is sufficiently well behaved, the discrete set of cumulants $W_{n,m}$ encode all of the information in the initial density, and any observable can be written as a function of this (infinite) set of arguments,
\begin{equation}
\label{func}
V_n = f(W_{n,m}) .
\end{equation}
The label $n$ in the cumulant $W_{n,m}$ specifies its rotational symmetry, which makes it simple to determine what combinations have the same rotational symmetry as $V_n$.  The label $m$ orders the modes in terms of the magnitude of $k$ as a Taylor series around $k=0$.  Modes with small $m$ represent small $k$, and encode large-scale structure.  Modes with large $m$ represent large $k$, and encode small-scale structure.  A reasonable hypothesis is that hydrodynamics is insensitive to very small-scale structure in the initial conditions (especially when damped by viscous effects), and the series can be truncated at some finite $m_{max}$.  That is, higher cumulants are not necessarily small, but the hydrodynamic response should nevertheless be insensitive to them. It is only after this truncation that this cumulant expansion becomes useful, rendering Eq.~\eqref{func} a function of a finite set of variables. 

For reference, we list the first few cumulants, neglecting overall normalization.
\begin{align*}
W_{1,1} &\propto  \left\{ r e^{i\phi}\right\}\\
W_{0,2} &\propto \left\{ r^2 \right\} 
- \left\{ re^{i\phi} \right\} \left\{ re^{-i\phi} \right\} \\
W_{2,2} &\propto  \left\{ r^2 e^{i2\phi} \right\} - \left\{ r e^{i\phi}\right\}^2 \\
W_{1,3} &\propto \left\{ r^3 e^{i\phi} \right\} - 2 \left\{ r e^{-i\phi} \right\} \left( \left\{ r^2 e^{i2\phi} \right\} - \left\{ r e^{i\phi} \right\}^2 \right)\\
& -\left\{ r e^{i\phi} \right\} \left( \left\{ r^2 \right\} - \left\{ r e^{i\phi} \right\} \left\{ r e^{-i\phi} \right\}  \right) -  \left\{ r e^{i\phi} \right\}^2 \left\{ r e^{-i\phi} \right\}\\
W_{3,3} &\propto \left\{ r^3 e^{i3\phi} \right\} + \frac 3 2 \left\{ r e^{i\phi} \right\} \left ( \left\{ r^2 e^{i2\phi} \right\} - \left\{ r e^{i\phi} \right\}^2 \right)\\
W_{4,4} &\propto \left\{ r^4 e^{i4\phi} \right\} - 2 \left\{ r^3 e^{i3\phi} \right\} \left\{ r e^{i\phi} \right\} - 3 \left\{ r^2 e^{i2\phi} \right\}^2  \\
&- 9 \left\{ r^2 e^{i2\phi} \right\} \left\{ r e^{i\phi} \right\}^2 + 11 \left\{ r e^{i\phi} \right\}^4 
,
\end{align*}
where the curly brackets represent a spatial average weighted by the density $\rho$,
\begin{equation}
\left\{\ldots\right\} \equiv \frac 
{\int d^2x \rho({\bf x}) \ldots}
{\int d^2x \rho({\bf x})} .
\end{equation}

One should note that taking the logarithm in the definition of cumulants was a key step.  As a result, all cumulants except $W_{1,1}$ are translationally invariant, making them suitable for predicting the final (translationally-invariant) momentum spectrum of particles.  It is for this same reason we can say that the index $m$ generally designates large- and small-scale structures in the initial density, rather than the structure at large and small radius with respect to some origin of coordinates.

Note also that the expressions simplify enormously if one shifts to a coordinate system such that $W_{1,1}$ = 0, or equivalently $\{x\} = \{y\} = 0$.  Such a centering shift is often done to simplify calculations, and will be used in the following.  However, it should be understood that these still represent translation-invariant quantities.

The next step is to notice that the azimuthally asymmetric cumulants, $W_{n,m}$ with $n\neq 0$, are typically small compared to the relevant scales in the problem in realistic events, such as the system size.  Therefore, we can write an estimator for Eq.~\eqref{func} as a perturbative series in powers of azimuthally asymmetric cumulants.  
Explicitly, all possible terms to second order can be written as
\begin{align}
\label{leadingorder}
V_{est,n} =&\sum_{m=n}^{m_{max}} k_{n,m} W_{n,m} \nonumber \\
& + \sum_{l=1}^{m_{max}} \sum_{m=l}^{m_{max}}\sum_{m' = |n-l|}^{m_{max}} K_{l,m,m'}  W_{l,m} W_{n-l,m'} \nonumber \\
& + O(W^3) .
\end{align}
It should be understood that only independent terms are kept---e.g., terms in the second sum with $n=l$ only have a linear dependence on anisotropic cumulants and should be combined with the relevant linear term.

Note that the validity of this expansion does not necessarily imply that the system evolution is well described by linearized hydrodynamics (even if a single linear term in Eq.~\eqref{leadingorder} is sufficient to describe a particular flow harmonic $V_n$), nor that the initial conditions can be well described as a small perturbation around a smooth and symmetric average.

Since $V_n$ is dimensionless, each of these terms should also be dimensionless; the (real) coefficients $k$ and $K$ must depend on the relevant scales of the system in order to render each term dimensionless.  

If this systematic expansion is valid,  a low-order truncation should give a good description of a full hydrodynamic simulation, and such an estimator can be further improved by increasing $m_{max}$ and/or adding higher terms in the power expansion, as well as adding terms that represent aspects that are missing in this treatment --- initial transverse flow and initial shear and bulk stress, as well as deviations from boost-invariance.  We will investigate which of these corrections is most important.

\section{Hydrodynamic calculation}
%
We calculate the flow harmonics Eq.~\eqref{vn} over the range of transverse momentum $0\leq p_T\leq 5$ GeV
using the boost-invariant  Lagrangian viscous relativistic hydrodynamical code v-USPhydro \cite{Noronha-Hostler:2013gga,Noronha-Hostler:2014dqa} with vanishing baryon chemical potential. v-USPhydro uses the simplest equations for the bulk scalar and the shear stress tensor in order to minimize the effect of the uncertainly regarding the unknown transport coefficients.  Here we can run ideal, shear, bulk, and shear+bulk separately in order to test each individual hydrodynamical expansion's affect on the prediction of the flow harmonics from the eccentricities, and to complement previous results using ideal hydrodynamics \cite{Gardim:2011xv} and the inclusion of shear viscosity \cite{Niemi:2012aj}.


As in \cite{Noronha-Hostler:2014dqa}, we consider a temperature dependent shear viscosity $\eta/s$ from the extended mass spectrum in the hadron gas phase \cite{NoronhaHostler:2008ju,NoronhaHostler:2012ug} and from lattice calculations in the high temperature region \cite{Nakamura:2004sy}, which was parameterized in \cite{Niemi:2012ry}. The $\zeta/s$ is inspired from the Buchel formula \cite{Buchel:2007mf}  and is the same as shown in \cite{Noronha-Hostler:2013gga,Noronha-Hostler:2014dqa}.  
The shear relaxation time is taken from \cite{Denicol:2010xn,Denicol:2011fa} and the bulk relaxation time is from \cite{Huang:2010sa}.

The initial conditions are calculated within a Monte Carlo Glauber model \cite{ic} at RHIC Au+Au $\sqrt{s_{NN}}$. The centrality class binning is the same as discussed in \cite{Noronha-Hostler:2014dqa} and for each centrality class we consider 150 events.  
In previous work \cite{Gardim:2011xv}, calculations were made with realistic fluctuating initial conditions --- specifically involving non-trivial, fluctuating dependence on rapidity and non-zero fluctuating initial transverse flow.
Here, we simplify by assuming that there is no initial transverse flow velocity $u^{x} = u^y =0$ and that there are no fluctuations in the rapidity (exact boost invariance in each event).  Additionally, when viscous corrections are considered we assume that both the bulk pressure, $\Pi$, and the shear stress tensor, $\pi^{\mu\nu}$ vanish at $\tau_0$.    Besides simplifying the calculations and allowing for a more clean study of the effect of, e.g., bulk viscosity, this also allows us to study the effect of these commonly used approximations by comparing to previous results. 

Full viscous corrections are considered within the Cooper-Frye freezeout (here we take $T_{FO}=150$ MeV) as detailed in \cite{Noronha-Hostler:2013gga,Noronha-Hostler:2014dqa} for the $\delta f$ of shear and bulk viscosity. In this paper we consider only the moments method for the bulk viscous corrections, however, as it was shown in \cite{Noronha-Hostler:2014dqa}, there is almost no difference between the final flow harmonics for the integrated $v_n$'s regardless of our choice in $\delta f$ for small $\zeta/s$.  Additionally, unlike in Eq. (C2-C3) in \cite{Noronha-Hostler:2013gga}, we do not use a $p_T$ weight to calculate the event plane vectors bur rather according to the definition \eqref{vn}.

In this paper v-USPhydro is not coupled to a hadronic afterburner, rather we fix the total number of $\pi^{+}$'s to 123, which is further explained in \cite{Noronha-Hostler:2013gga}. We would not expect that the hadronic decays would play a large role here but we leave this to be tested in a future work. 

Furthermore, we compare our results to our previous NeXSPheRIO results in \cite{Gardim:2011xv}. NeXSPheRIO provides a good description of rapidity and transverse momentum spectra \cite{Qian:2007}, elliptic flow \cite{Andrade:2008,Gardim:2012}, rapidity-even directed flow at midrapidity \cite{Gardim:2011qn}, triangular and quadrangular flow  \cite{Gardim:2012}. In addition, it  reproduces the long-range structures observed in two-particle correlations \cite{Takahashi:2009na} and  the breaking of factorization seen in two-particle correlations \cite{Gardim:2013}. 

NeXSPHeRIO is a 3+1  relativistic hydrodynamical code which solves the equations of ideal relativistic hydrodynamics using fluctuating initial conditions from the event generator NeXus \cite{Drescher:2001}. 

NeXus aims at a realistic and consistent approach of the initial stage of nuclear collisions. It is a Monte-Carlo generator which takes into account not only the fluctuations of nucleon positions within nuclei as above, but also fluctuations at the partonic level: the momentum of each nucleon is shared between one or several participants and remnants, which implies non-trivial dynamical fluctuations in each nucleon-nucleon collisions. The resulting full energy-momentum tensor is matched to a hydrodynamic form, resulting in a fluctuating flow field in addition to a fluctuating initial energy density, in all three spatial dimensions, with the transverse length scale of the fluctuations set mostly by the size of the incoming nucleons.

150 NeXus events were generated \cite{Gardim:2011xv} in each of the 10 \% centrality classes
studied and the equations of ideal hydrodynamics were solved for each event. The end of the hydrodynamic evolution is assumed to occur at a constant freeze-out temperature for each event and discrete particles are emitted using a Monte-Carlo generator (resonances decays are implemented.) For each hydro event, we run the Monte-Carlo generator many times, so that we can do the flow analysis using approximately $6 \times 10^5$ particles per event. This significantly reduces statistical noise and allows for an accurate determination of $v_n$ and $\Psi_n$ in every event. 
(It also suppresses nonflow correlations from, e.g., particle decays.) These quantities are then calculated by Eq. (\ref{vn}), considering  all particles in the pseudorapidity interval $ |\eta| < 1$.
\section{Results}
\subsection{$v_2$}
Consider an estimator, Eq.~\eqref{leadingorder}, for $V_2$ with $m_{max} = 2$.  This gives
\begin{equation}
V_{est,2} = k_{2,2} W_{2,2} + O(W^3) ,
\end{equation}
The next term in the power series would be proportional to $(W_{2,2})^3$, and the next linear term, describing smaller scale structure, proportional to $W_{2,4}$.

One of the most natural scales to compare to is the global system size, or $W_{0,2} \propto \{r^2\}$.  
Choosing $k_{2,2} = k'/W_{0,2}$, with $k'$ a dimensionless constant, we finally obtain the famous participant eccentricity $\varepsilon_2$ and participant plane $\Phi_2$, as the magnitude and phase of the ratio of the lowest possible cumulants with the same symmetries as the elliptic flow vector $V_2$
%
%
\begin{equation}
\label{eps2}
\varepsilon_2 e^{i2\Phi_2} \equiv -2 \frac {W_{2,2}} {W_{0,2}} = - \frac {\{r^2 e^{i2\phi}\}} {\{r^2\}} .
\end{equation}
%
Recall that the expression on the right has been simplified by assuming a centered coordinate system, but it actually represents a translation invariant quantity.

This has long been thought to be a reasonably good estimator
\begin{equation}
\label{v2est}
V_2 \left(= v_2 e^{i2\Psi_2}\right) \simeq V_{est,2} \simeq  k\ \varepsilon_2 e^{i2\Phi_2} 
\end{equation}
The minus sign in Eq.~\eqref{eps2} is arbitrary, but is used in anticipation of an expected positive value for the coefficient $k$ --- the \textit{minor} axis of an elliptical initial density corresponds to the direction of steepest gradient, and therefore the direction that more particles are pushed.  

We can actually test this idea quantitatively by calculating a large set of events and computing the quality estimator Eq.~\eqref{Q}, which in this case becomes explicitly
\begin{equation}
\label{Qv2}
Q_2 = \frac 
{\left\langle v_2 \varepsilon_2 \cos 2(\Psi_2 - \Phi_2) \right\rangle}
{\sqrt{\left\langle v_2^2\right\rangle \left\langle \varepsilon_2^2\right\rangle}} .
\end{equation}
\begin{figure}
\includegraphics[width=\linewidth]{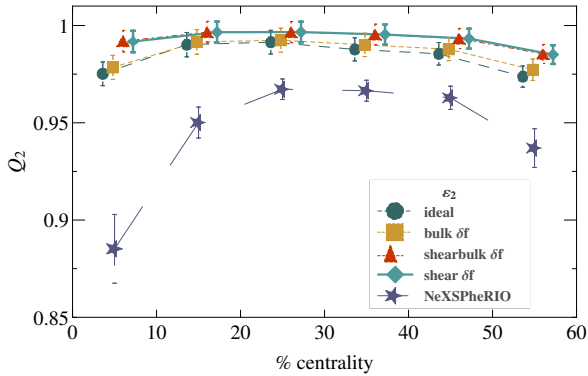}
\caption{(Color online) Measure of the quality $Q_2$ \eqref{Q} of the elliptic flow estimator defined in Eq.~\eqref{v2est}; i.e., the event-by-event linear correlation coefficient \eqref{Qv2} between the elliptic flow and initial eccentricity vectors over 150 events in each centrality bin.  Computations were performed with ideal hydrodynamics (``ideal''), as well as including non-zero shear viscosity (``shear''), non-zero bulk viscosity (``bulk''), or both (``shearbulk''), using v-USPhydro. Previous results \cite{Gardim:2011xv} from NeXSPheRIO are also shown (star). The points have been given an x-offset for readability.}
\label{v2}
\end{figure}
This was done for the first time in Ref.~\cite{Gardim:2011xv}.  Equation \eqref{Qv2} was calculated using both energy density and entropy density, with the result for energy density reproduced here in Fig.~\ref{v2}.
In both cases, Eq.~\eqref{v2est} is indeed a good approximation as a vector equation --- the event plane $\Psi_2$ is approximately the same as the participant plane $\Phi_2$ in each event, and the elliptic flow coefficient $v_2$ is proportional to the eccentricity $\varepsilon_2$.  

In this approximation, the coefficient $k$ contains all relevant information about fluid properties such as viscosity, as well as freeze out and subsequent evolution of the system, and this is the same in every collision event at a given centrality.  Conversely the only thing that changes from one event to the next is the initial condition, and the only aspect that matters is the large scale structure --- in particular the only allowed combinations of the lowest cumulants ($W_{2,2}$ and $W_{0,2}$) to quadratic order.

Unsurprisingly, the simple linear dependence of Eq.\ (\ref{v2est}) allows for straightforward relationships to be drawn between the initial state and various observables. 
Indeed, one can find combinations of observables that effectively remove the dependence on medium properties \cite{Bhalerao:2011yg}, or conversely, to isolate medium properties, in a way that would not be possible without this knowledge \cite{Luzum:2012wu}.
Note that, in principle, one can use any choice of basis to characterize the initial transverse density \cite{ColemanSmith:2012ka, Floerchinger:2013rya}.  However, one benefits enormously by choosing a basis that converges most quickly (in the sense of the expansion Eq.~\eqref{leadingorder}).  The fewer number of terms that can describe the hydrodynamic response to the desired accuracy, the more useful and powerful is the resulting relation.  Even at linear order, if more than just the lowest term in the $m$ expansion is required, the expression for the magnitude $v_n$ and $\Psi_n$ become much more complicated.

Indeed, it was shown in Ref.~\cite{Gardim:2011xv} that adding the next term (i.e., considering a linear combination of $W_{2,2}$ and $W_{2,4}$) has a negligible effect, indicating that the small deviation of $Q_2$ from 1 is due either to nonlinear terms, or from the initial transverse flow and rapidity-dependent fluctuations.   If we compare the previous (NeXSPheRIO ideal hydrodynamics) results to the ideal hydrodynamic result in this work, as seen in Fig.~\ref{v2}, we can see that the quality $Q_2$ becomes extremely close to 1.  Thus, rather than nonlinear terms, most of the deviation from 1 in the previous results is explained by the latter effects, and they appear to be most important for central collisions.  We save a more thorough investigation of these effects for future work.

However, the previous result may not be entirely realistic because it neglects viscous effects.  The later results of Ref.~\cite{Niemi:2012aj} imply that shear viscosity should improve the estimator.   We confirm this result, as can be seen in Fig.~\ref{v2}.  Adding bulk viscosity to ideal hydrodynamics has a similar effect.  With the value of bulk viscosity used here, the effect is smaller, and when bulk viscosity is included in addition to shear viscosity there is a negligible difference in $Q$.  Thus, the previous results from NeXSPheRIO represent something of a worst-case scenario, and reality will likely lie somewhere in between these results.

So in general, remarkably, one can predict quite accurately $v_2$ in a given event by only knowing $\varepsilon_2$, regardless of any other details of the initial condition.
%
%
%
\subsection{$v_3$}
\begin{figure}
\includegraphics[width=\linewidth]{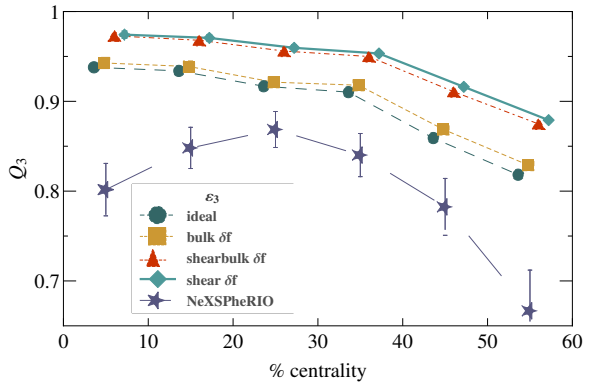}
\caption{(Color online) Measure of the quality $Q_3$ \eqref{Qv3} of initial triangularity as an estimator of triangular flow, Eq. \eqref{v3est}, for ideal and several kinds of viscous hydrodynamics using v-USPhydro. NeXSPheRIO results are also shown (star) \cite{Gardim:2011xv}. The points have been given an x-offset for readability.}
\label{v3}
\end{figure}
%
%
%
The lowest cumulant with the correct symmetry for $V_3$ is $W_{3,3}$.  The only rotationally-invariant cumulant with $m\leq 3$ that can provide a scale for creating a dimensionless ratio is $W_{0,2}$.  Thus one possible choice for estimator is
\begin{equation}
-\frac {W_{3,3}} {W_{0,2}} \propto -\frac {\{r^3 e^{i3\phi}\}} {\{r^2\}^{3/2}} .
\end{equation}
A more popular choice for scale is $\{r^3\}$, which results in the standard `triangularity'
\begin{equation}
\label{eps3}
\varepsilon_3 e^{i3\Phi_3} \equiv - \frac {\{r^3 e^{i3\phi}\}} {\{r^3\}} 
\end{equation}
Note that the quantity $\{r^3\}$ is not actually a cumulant, nor a simple function of cumulants.  However, it was found that both of the above estimators are equally  good \cite{Gardim:2011xv}, and so the distinction is not important.  For reference, in the latter case we have explicitly
\begin{equation}
\label{v3est}
V_{est,3} = k\ \varepsilon_3 e^{i3\Phi_3}
\end{equation}
and Eq.~\eqref{Q} becomes
\begin{equation}
\label{Qv3}
Q_3 = \frac 
{\left\langle v_3 \varepsilon_3 \cos 3(\Psi_3 - \Phi_3) \right\rangle}
{\sqrt{\left\langle v_3^2\right\rangle \left\langle \varepsilon_3^2\right\rangle}} .
\end{equation}
The results for this estimator are shown in Fig.~\ref{v3}.  Similar to the elliptic flow results, both shear viscosity and bulk viscosity improve the quality $Q_3$ of the estimator compared to the ideal hydrodynamics results.  However, adding bulk viscosity to a calculation with non-zero shear viscosity results in a slight decrease of $Q_3$.  This can be understood as follows.

\begin{figure}
\includegraphics[width=\linewidth]{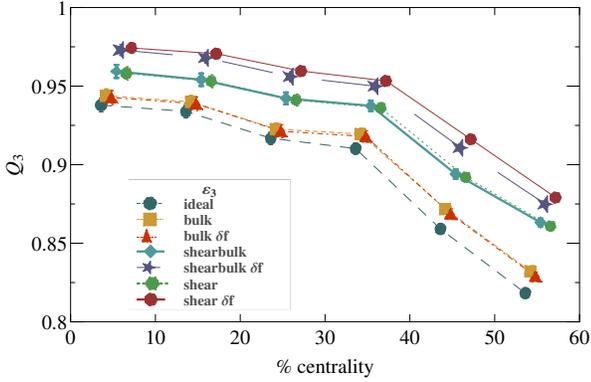}
\caption{(Color online) Measure of the quality $Q_3$ \eqref{Qv3} of initial triangularity as an estimator Eq. \eqref{v3est}, for ideal and viscous hydrodynamics with and without the viscous correction to the equilibrium distribution function $\delta f$ using v-USPhydro. The points have been given an x-offset for readability.}
\label{df}
\end{figure}

Bulk viscosity, like shear viscosity, has a smoothing effect on the hydrodynamic evolution
 that decreases the dependence on higher cumulants
 (i.e., on small-scale structure in the initial condition).  This can be seen in Fig.~\ref{df} by comparing results with and without shear and bulk viscosity, but neglecting the viscous correction to the distribution function --- i.e., assuming an isotropic, thermal particle distribution function at freeze out.  However, whenever viscosity is present, there is necessarily a \textit{local} anisotropy, represented by a correction $\delta f$.   This correction is a complicated nonlinear function that does not necessarily follow the same intuition as the hydrodynamic equations --- that viscosity damps small-scale structure.  
 
 Indeed, we see that adding $\delta f$ in the case of shear viscosity increases the linear correlation between triangularity and triangular flow $Q_3$, while for bulk viscosity, adding $\delta f$ decreases $Q_3$.  So shear viscosity appears to always increase $Q_3$, while bulk viscosity displays an interplay between its effect on the global hydrodynamic evolution and the local distribution function whose combination can have a different net effect on $Q_3$ depending on the situation.
 
One can add corrections to this linear relation either by keeping $m_{max} = 3$ and going to quadratic order, which gives the possible term $W_{1,3}W_{2,2}$,  or by adding the next highest linear term, $W_{3,5}$.  

Following the traditional way of normalizing these terms to create dimensionless quantities, we can define the a two-index anisotropy coefficients and angles
\begin{equation}
\varepsilon_{n,m} e^{i\Phi_{n,m}} \equiv - \frac {\{r^m e^{in\phi}\}} {\{r^m\}} .
\end{equation}
%
Note the ordering of $m$ and $n$, which we choose here following the convention for $W_{n,m}$, originally from Ref.~\cite{Teaney:2010vd}, but which is opposite of the $\varepsilon_{m,n}$ from Ref.~\cite{Gardim:2011xv}. The order should always be clear since valid cumulants always have $m\geq n$. 

Anticipating our discussion of $v_1$, we can define the lowest translation-invariant anisotropy coefficient as $\varepsilon_1 = \varepsilon_{1,3}$ and $\Phi_1 = \Phi_{1,3}$.  (Note that $W_{1,1}$ simply indicates the location of the center of the system, and such a translation-variant quantity cannot be expected to predict any aspect of any $V_n$.  As such, the lowest relevant cumulant in the first harmonic is $W_{1,3}$). 

Adding the corrections then gives the following full estimator
\begin{equation}
\label{v3:allterms}
V_{est,3} = k\ \varepsilon_3 e^{i3\Phi_3} + k'\ \varepsilon_{3,5} e^{i3\Phi_{3,5}} + k''\ \varepsilon_1 \varepsilon_2 e^{i(\Phi_1 + 2\Phi_2)} .
\end{equation}

If two or more of the coefficients $k$ are allowed to be non-zero, the quality measure $Q_3$ now becomes a more complicated function.  The coefficients of the best estimator are given by the solution of a set of linear equations, Eq.~\eqref{linearequations}, and are then plugged into Eq.~\eqref{Q} to obtain $Q_3$.


\begin{figure}
\includegraphics[width=\linewidth]{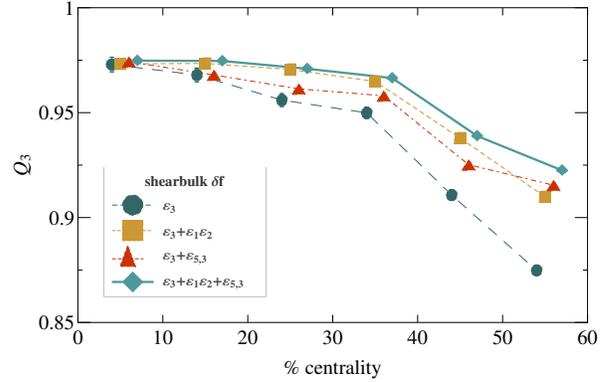}
\caption{(Color online) Measure of the quality $Q_3$ \eqref{Q} of improved estimators Eq.~\eqref{v3:allterms} in viscous hydrodynamics using v-USPhydro including both shear and bulk viscosity. The points have been given an x-offset for readability.}
\label{v3new}
\end{figure}

In Fig.~\ref{v3new}, we show the effect of adding the corrections individually and together.  One can see that the higher cumulant correction $\varepsilon_{3,5}$, representing smaller-scale structure, has a negligible effect in central collisions, with a small but increasing importance for more peripheral collisions.  The non-linear term $\varepsilon_1\varepsilon_2$ is also negligible in central collisions, but is more important than the higher-order cumulant in non-central collisions. 

In general, the systematic framework described in Sec.~\ref{expansion} appears to be indeed valid --- the lowest linear term is by far the most important and can predict $v_3$ to very good precision, while systematically adding terms in the expansion of Eq.~\eqref{leadingorder} improves the estimator even more.   Non-linear terms (in particular those involving the second harmonic, as discussed below) appear to be more important than small-scale structure in the initial conditions, as quantified by higher-order cumulants.  However, demanding a linear correlation much larger than $\sim$95\% may require the inclusion of a quickly increasing number of terms, and will be complicated further by the presence of non-trivial rapidity dependence and initial flow in a more realistic simulation.
\subsection{$v_4$ and $v_5$}
Equation \eqref{leadingorder} for $V_4$ allows for a linear term, but requires a cumulant with $m=4$.  Conversely, if one demands $m_{max} = 2$, allowing only the lowest cumulants, then a quadratic term is required \cite{Borghini:2005kd}.  Explicitly we have:
\begin{align}
V_{est,4} &= k_{4,4} W_{4,4} + K_{2,2,2} W_{2,2} W_{2,2} + O(W^2) ,
\end{align}
where the next omitted terms are quadratic with $m_{max}=4$ [see Eq. (\ref{eqn:allv4})].
It turns out that both these terms are important---the quadratic term being particularly important not because the hydrodynamic response $K_{2,2,2}$ to the larger scale structure ($m=2$ compared to $m=4$) is strong enough for the quadratic term to overwhelm a much larger linear cumulant, but because it is typically just as large.  That is, the elliptical overlap region between nuclei colliding at non-zero impact parameter causes modes with azimuthal harmonic $n=2$ to be systematically larger than other modes.  Thus, we must use a separate power counting for these modes, since the quadratic term $(W_{2,2})^2$ can be the same size as $W_{4,4}$ in non-central collisions.  In central collisions, on the other hand, the quadratic term is not important \cite{Gardim:2011xv}.

In a centered coordinate system, this relation can be rearranged to read
\begin{equation}
\label{v4:2terms}
V_{est,4}= \tilde{k} \{r^4 e^{i4\phi}\} + \tilde{k'} \{r^2 e^{i2\phi}\}^2 .
\end{equation}
(One should note that the first term here does not correspond to $W_{2,2}$, but rather a linear combination of the terms).
Thus, if we define a ``quadrangularity'' in analogy with eccentricity  and triangularity
\begin{equation}
\varepsilon_4 e^{i4\Phi_4} \equiv - \frac {\{r^4 e^{i4\phi}\}} {\{r^4\}} ,
\end{equation}
we can obtain the canonical estimator which has been shown to accurately predict quadrangular flow event-by-event \cite{Gardim:2011xv}
\begin{equation}\label{eqn:canestv4}
V_4 \simeq V_{est,4} = k\ \varepsilon_4 e^{i4\Phi_4} + k' \varepsilon_2^2 e^{i4\Phi_2} .
\end{equation}
The vector nature of the flow coefficients becomes especially important here.  Because this relation has two vector terms, it is more complicated than the simple relations that worked for $v_2$ and $v_3$.  For example, if this relation holds, the magnitude $v_4$ is not a function only of the magnitudes $\varepsilon_4$ and $\varepsilon_2^2$, but instead depends on both the magnitude and direction of the anisotropy vectors
\begin{equation}
v_4^2 = |V_4|^2 = k^2 \varepsilon_4^2 + k'^2 \varepsilon_2^4 + 2 k k' \varepsilon_4\varepsilon_2^2 \cos 4(\Phi_2 - \Phi_4) .
\end{equation}
The coefficients of the best estimator $k$ and $k'$ are again given by the solution of 2 linear equations, Eq.~\eqref{linearequations}, which are then plugged into Eq.~\eqref{Q} to find $Q_4$.

The same discussion applies for the fifth azimuthal harmonic.  A good estimator requires both a linear contribution $W_{5,5}$ and a quadratic term $W_{2,2} W_{3,3}$, or
\begin{equation}
\label{v5:2terms}
V_5 \simeq V_{est,5} = k\ \varepsilon_5 e^{i5\Phi_5} + k' \varepsilon_2\varepsilon_3 e^{i(2\Phi_2 + 3\Phi_3)} , 
\end{equation}
with
\begin{equation}
\varepsilon_5 e^{i5\Phi_5} \equiv - \frac {\{r^5 e^{i5\phi}\}} {\{r^5\}} .
\end{equation}
\begin{figure}
\includegraphics[width=\linewidth]{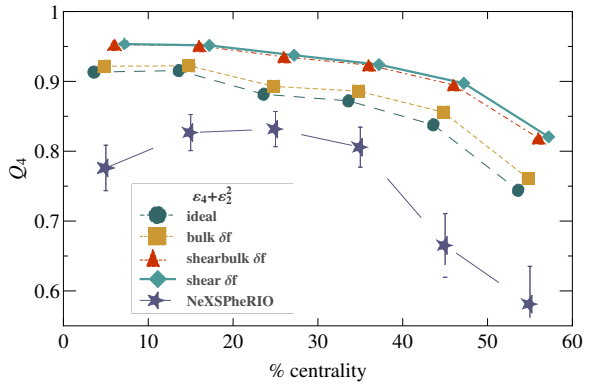}
\caption{(Color online) Measure of the quality $Q_4$ \eqref{Q} of the two-term estimator Eq.~\eqref{eqn:canestv4}, for ideal and several kinds of viscous hydrodynamics using v-USPhydro. NeXSPheRIO results are also shown (star). The points have been given an x-offset for readability.}
\label{v4}
\end{figure}

\begin{figure}
\includegraphics[width=\linewidth]{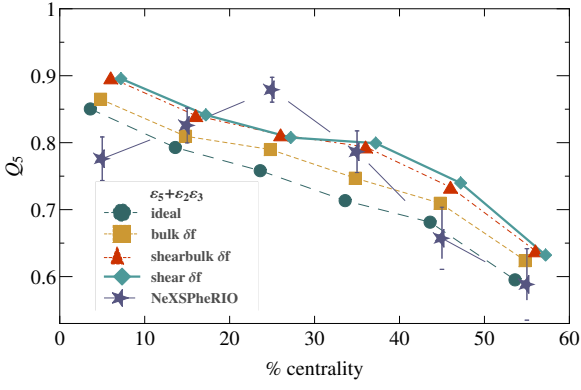}
\caption{(Color online) Measure of the quality $Q_5$ \eqref{Q} of the two-term estimator Eq.~\eqref{v5:2terms}, for ideal and several kinds of viscous hydrodynamics using v-USPhydro. NeXSPheRIO results are also shown (star). The points have been given an x-offset for readability.}
\label{v5}
\end{figure}

Our results are shown in Figs.~\ref{v4} and \ref{v5}, respectively.  They show similar behavior as the lower harmonics.  Shear viscosity increases the quality of the estimator, while the addition of bulk improves the ideal hydrodynamic result, but not the shear viscous result.

We can also investigate corrections to this estimator.  Equation \eqref{leadingorder} for $V_4$, with $m_{max}=4$ reads in full
\begin{align}
\label{v4cum}
V_{est,4} &= k_{4,4} W_{4,4} + K_{2,2,2} W_{2,2}W_{2,2} \nonumber\\
& + K_{2,4,2} W_{2,4}W_{2,2} + K_{2,4,4} W_{2,4}W_{2,4} \nonumber\\
& + K_{1,3,3} W_{1,3}W_{3,3} + O(W^2) ,
\end{align}
and we can therefore propose a potentially improved estimator 
\begin{align}
\label{eqn:allv4}
V_{est,4} &= k \varepsilon_{4}e^{i4\Phi_4}+k'\varepsilon^2_2e^{i4\Phi_2}+k''\varepsilon_2\varepsilon_{2,4}e^{i2(\Phi_2+
\Phi_{2,4})}\nonumber\\
&+k'''\varepsilon_{2,4}^2e^{i4\Phi_{2,4}}+k''''\varepsilon_1\varepsilon_3e^{i(\Phi_1+3\Phi_3)} .
\end{align}

Similarly, we can write a full estimator for $V_5$ up to $m_{max}=5$:
\begin{align}
\label{eqn:allv5}
V_{est,5} &= k \varepsilon_{5}e^{i4\Phi_5}+k'\varepsilon_3\varepsilon_2e^{i(3\Phi_3+2\Phi_2})\nonumber\\
&+k''\varepsilon_{3,5}\varepsilon_2e^{i(3\Phi_{3,5}+2\Phi_2)}+k'''\varepsilon_{3}\varepsilon_{2,4}e^{i(3
\Phi_{3}+2\Phi_{2,4})} \nonumber\\
&+ k''''\varepsilon_{3,5}\varepsilon_{2,4}e^{i(3\Phi_{3,5}+2\Phi_{2,4})}
+k'''''\varepsilon_1\varepsilon_4e^{i(\Phi_1+4\Phi_4)}  \nonumber\\
&+ k''''''\varepsilon_{1,5}\varepsilon_4e^{i(\Phi_{1,5}+4\Phi_4)}.
\end{align}

\begin{figure}
\includegraphics[width=\linewidth]{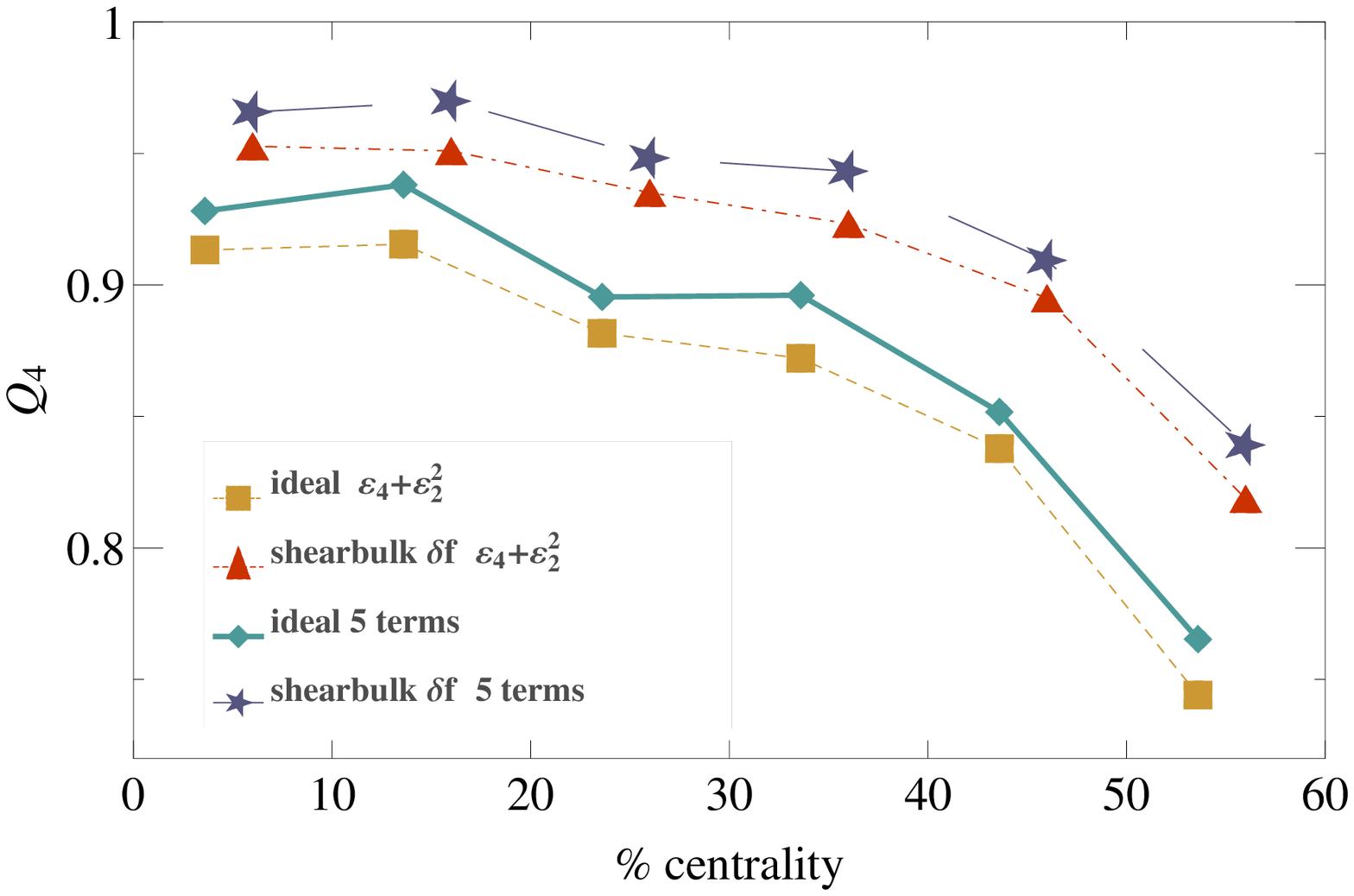}
\caption{(Color online) Measure of the quality $Q_4$ \eqref{Q} for the two-term estimator Eq.~\eqref{v4:2terms} and improved estimator Eq.~\eqref{eqn:allv4}, for ideal and shear+bulk $\delta f$ viscous hydrodynamics using v-USPhydro. The points have been given an x-offset for readability.}
\label{v4new}
\end{figure}

\begin{figure}
\includegraphics[width=\linewidth]{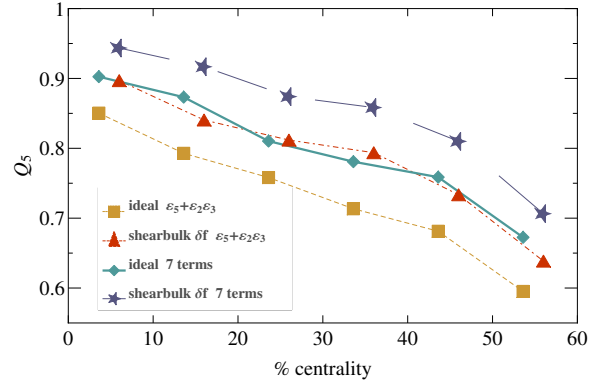}
\caption{(Color online) Measure of the quality $Q_5$ \eqref{Q} for the two-term estimator Eq.~\eqref{v5:2terms} and improved estimator Eq.~\eqref{eqn:allv5}, for ideal and shear+bulk $\delta f$ viscous hydrodynamics using v-USPhydro. The points have been given an x-offset for readability.}
\label{v5new}
\end{figure}

In Figs.~\ref{v4new} and ~\ref{v5new}, one can see that the corrections do indeed improve the estimator, though only a small amount compared to the first two terms alone, which already proved a good estimator.
\subsection{$v_1$}
The lowest translationally-invariant cumulant with the same symmetries as $V_1$ is $W_{1,3}$. 
Similar to the other harmonics, a dimensionless asymmetry parameter can be defined from this, called the ``dipole asymmetry'' \cite{Teaney:2010vd}
\begin{equation}
\varepsilon_1 e^{i\Phi_1} \equiv - \frac {\{r^3 e^{i\phi}\}} {\{r^3\}} .
\end{equation}
Directed flow has been less studied than $v_2$--$v_5$ --- it appeared to be reasonably well correlated with the dipole asymmetry $\varepsilon_1$, but this has not been quantified.  Nor have the effects of corrections to the simplest linear relationship been studied.  

%
The next order linear correction, then, with $m_{max}=5$, can be investigated by adding a term proportional to $\varepsilon_{1,5}$.

Similarly to $v_4$ and $v_5$, there are possible quadratic terms involving the second harmonic that may be important -- in this case two terms, $W_{1,3}^*W_{2,2}$ and $W_{3,3}W_{2,2}^*$, have the correct symmetries (although the latter is typically larger than the former).   
%

We test the importance of each of these by using the estimator
\begin{eqnarray}\label{V1est}
V_1 \simeq V_{est,1}&=&k \varepsilon_1 e^{i\Phi_1}+ k' \varepsilon_2\varepsilon_3 e^{i(3\Phi_3-2\Phi_2)} + k'' \varepsilon_{1,5}e^{i\Phi_{1,5}} \nonumber\\
&+& k''' \varepsilon_2\varepsilon_1 e^{i(2\Phi_2-\Phi_1)},
\end{eqnarray}
%
with various of the coefficients set to zero to study the importance of the inclusion of each term.

\begin{figure}
\includegraphics[width=\linewidth]{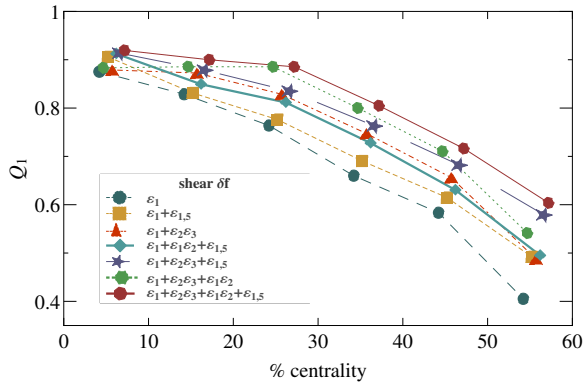}
\caption{(Color online) $Q_1$ \eqref{Q} for various combinations of terms in estimator \eqref{V1est}.  The hydrodynamic calculation includes shear viscosity and the standard correction to the distribution function $\delta$f. 
The points have been given an x-offset for readability.}
\label{v1ho}
\end{figure}

In Fig.\ \ref{v1ho}, various combinations are compared for $v_1$ calculated with shear viscosity included.  The lowest order cumulant, $\varepsilon_1$, is a very good estimator for central collisions, but is poorer in peripheral collisions.  In the case of $v_4$ and $v_5$ a similar (but stronger) degradation of the correlation with $\varepsilon_n$ indicated the importance of quadratic terms \cite{Gardim:2011xv}.  
In Fig.\ \ref{v1ho}, we see that the nonlinear terms do indeed have an effect.  In particular, adding the $\varepsilon_2\varepsilon_3$ term significantly improves the quality of the estimator, while the $\varepsilon_2\varepsilon_1$ term appears to be much less important.  However, even with all three terms, the quality still does not exceed $Q_1=60\%$ in peripheral collisions.

Directed flow, $v_1$, also has a non-negligible dependence on higher cumulants, and therefore smaller scale structure in the initial density.  Adding $\varepsilon_{1,5}$ provides a non-negligible improvement to the estimator, even in the presence of viscosity.  This is in stark contrast to $v_2$, which showed a negligible dependence on higher cumulants, even in the case of ideal hydrodynamics \cite{Gardim:2011xv}.  So both higher cumulants and nonlinear terms are important for non-central collisions, and further corrections will be necessary to get $Q_n>$ 0.8--0.9 at all centralities.

\begin{figure}
\includegraphics[width=\linewidth]{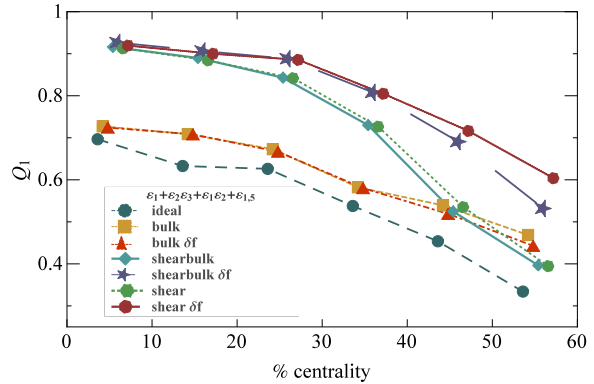}
\caption{(Color online) Measure of the quality $Q_1$ \eqref{Q} for $v_1$ computed 
using the  estimator given by Eq. \eqref{V1est}, for ideal and several kinds of viscous hydrodynamics using v-USPhydro. The points have been given an x-offset for readability.}
\label{v1pt1}
\end{figure}

In Fig.\ \ref{v1pt1} the results using Eq.\ (\ref{V1est}) are shown.   One can see that as with our previous results shear viscosity gives the highest quality estimator. However, shear and bulk together are almost identical to shear with the exclusion of the most peripheral collisions.  $v_1$ also appears to be the most strongly affected by shear (and shear+bulk) viscosity compared to the other flow harmonics.  One can see a large 
difference
between bulk and ideal vs $v_1$ when shear is present, which is in stark contrast to our results for $v_2$ where the various viscosities play only a minimal role.  
%
%
%
%
%
%
%
%
%
%
%

\begin{figure}
\includegraphics[width=\linewidth]{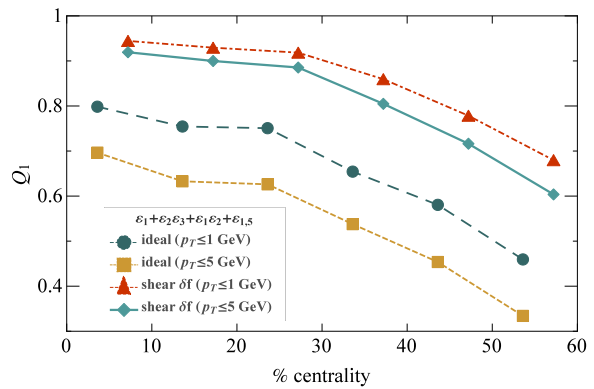}
\caption{(Color online) $Q_1$ \eqref{Q} of estimator \eqref{V1est} for $p_T$-integrated directed flow ($p_T<$ 5 GeV), compared to $Q_1$ for the directed flow of hadrons with $p_T<$ 1 GeV.  The hydrodynamic calculation includes shear viscosity and the standard correction to the distribution function $\delta$f.
The points have been given an x-offset for readability.}
\label{v1comp}
\end{figure}

Directed flow has a more complicated dependence on transverse momentum compared to the other harmonics.  Low $p_T$ hadrons tend to flow in the opposite direction compared to high $p_T$ hadrons.  I.e., $\Psi_1(p_T)$ changes by a phase of $\pi$ from low to high transverse momentum.  Alternately, if the flow vector is projected onto a fixed direction, as in common flow measurements, directed flow changes sign as a function of $p_T$.  
This non-trivial dependence also affects our ability to predict $v_1$ from the initial dipole asymmetry.  
In Fig. \ref{v1comp} we compare $Q_1$ for $p_T$-integrated flow to the same quantity calculated for directed flow with a smaller range $p_T\leq 1$ GeV.  It turns out that low-$p_T$ hadrons are better correlated with the initial dipole asymmetry, as well as the improved estimator \eqref{V1est}, compared to the set of all hadrons.   While the effect is not large, it is much larger for $v_1$ than for the other harmonics.  
We save a more detailed investigation into the $p_T$-dependence for future work.

\section{Conclusions}
In this paper we have tested the viscous hydrodynamical response to event-by-event initial conditions of the energy density profile.  In the most state of the art relativistic hydrodynamical calculations many factors contribute to the final anisotropic flow such as the initial flow, fluctuations in the initial energy density in rapidity, and transport coefficients.  In our calculations we find that the inclusion of shear viscosity within relativistic hydrodynamical calculations provides the cleanest relationship between the final flow harmonics and the large-scale structures in the initial density, represented by eccentricities $\varepsilon_n$.  Once bulk viscosity is included the sensitivity to other aspects of the initial conditions increases but the correlation between $\varepsilon_n$ and $v_n$ is still stronger than for ideal hydrodynamics.  

Furthermore, we compared the results using the ideal version of v-USPhydro  to that of NeXSPheRIO.  The primary differences between the two codes is that SPheRIO is a 3+1 code with NEXUS  initial conditions that have fluctuations in the rapidity direction and that NEXUS provides a non-zero, fluctuating initial flow.  One can see in Figs.\ \ref{v2}-\ref{v3}, \ref{v4}-\ref{v5} that generally the NeXSPheRIO results remain below the v-USPhydro results (with the exception of $v_5$).  It appears that the central collisions are the most sensitive to the differences between v-USPhydro and NeXSPheRIO, which indicates that either initial non-zero flow or fluctuations in the rapidity direction partially washes out the effects of the initial eccentricities on the final anisotropic flow.  It appears that the effect of initial flow and longitudinal fluctuations are then not as important for $v_5$ because NeXSPheRIO actually provides the same or better estimator than the ideal results from v-USPhydro (with the exception of the most central collisions).  Additionally, even for the other flow harmonics it appears that the mid-centrality flow harmonics (around $20-40\%$) are the least affected by the initial flow and longitudinal fluctuations, which indicates this is the region (combined as well as with $v_5$) most able to give us information needed to understand the initial conditions even in the most state of the art calculations. 

One question raised by these results is why the initial conditions from peripheral collisions are such poor estimators for the final flow harmonics.  In both $v_1$ and $v_5$ for the most peripheral collisions the initial eccentricities do not appear to be able to predict the final flow harmonics as well as for other harmonics or for other centralities.  It may well be that the peripheral collisions (especially for $v_1$ and $v_5$) are more dependent on the higher order cumulants of the initial energy density profile.  

\acknowledgments
This work was supported in part  by Funda\c{c}\~{a}o de Amparo \`{a} Pesquisa do Estado de S\~{a}o Paulo
(FAPESP), Conselho Nacional de Desenvolvimento Cient\'{\i}fico e Tecnol\'ogico 
 (CNPq) e NAP-QCD/Universidade de S\~ao Paulo, in Brazil.
M.L~was supported by the Office of Nuclear Physics in the US Department of Energy's Office of Science under Contract No. DE-AC02-05CH11231.  JNH acknowledges
support from the US-DOE Nuclear Science Grant No. DE-FG02-93ER40764.
%

\end{document}